\documentclass[a4paper,11pt]{article}
\pdfoutput=1 % if your are submitting a pdflatex (i.e. if you have
\usepackage{jinstpub} % for details on the use of the package, please
\usepackage[hang,tight]{subfigure}
\usepackage[normalem]{ulem}

\title{\boldmath Accuracy of the geometric-mean method for determining spatial resolutions of tracking detectors in the presence of multiple Coulomb scattering} 

\author{Aiwu Zhang}
\author{and Marcus Hohlmann}
\affiliation{Department of Physics and Space Sciences, Florida Institute of Technology, \\150 W. University Blvd, Melbourne FL 32901, USA}

\emailAdd{azhang@fit.edu}

\abstract{The geometric-mean method is often used to estimate the spatial resolution of a position-sensitive detector probed by tracks. It calculates the resolution solely from measured track data without using a detailed tracking simulation and without considering multiple Coulomb scattering effects. Two separate linear track fits are performed on the same data, one excluding and the other including the hit from the probed detector. The geometric mean of the widths of the corresponding exclusive and inclusive residual distributions for the probed detector is then taken as a measure of the intrinsic spatial resolution of the probed detector: $\sigma=\sqrt{\sigma_{ex}\cdot\sigma_{in}}$. The validity of this method is examined for a range of resolutions with a stand-alone Geant4 Monte Carlo simulation that specifically takes multiple Coulomb scattering in the tracking detector materials into account. Using simulated as well as actual tracking data from a representative beam test scenario, we find that the geometric-mean method gives systematically inaccurate spatial resolution results. Good resolutions are estimated as poor and vice versa. The more the resolutions of reference detectors and probed detector differ, the larger the systematic bias. An attempt to correct this inaccuracy by statistically subtracting multiple-scattering effects from geometric-mean results leads to resolutions that are typically too optimistic by 10-50\%. This supports an earlier critique of this method based on simulation studies that did not take multiple scattering into account.}


\begin{document}
\maketitle
\flushbottom

\section{Introduction} \label{sec:intro}
The geometric-mean method for determining the spatial resolution of a position-sensitive detector has become popular in studies of Micro-Pattern Gas Detectors (MPGDs) such as Gas Electron Multipliers (GEMs) and Micromegas detectors~\cite{NIMA-538-2005, NIMA-602-2009, NIMA-628-2011, NIMA-782-2015}. In this method, a detector is probed by charged-particle tracks that are reconstructed with a reference tracker in two separate ways by either excluding or including the hit of the probed detector in linear track fits. Corresponding ``exclusive'' and ``inclusive'' residual distributions are then calculated for the probed detector. The width $\sigma_{ex}$ of the exclusive residual distribution is typically larger than the true intrinsic resolution $\sigma$ of the probed detector because it convolutes the intrinsic resolution of the probed detector with the track error at the position of the probed detector. The width $\sigma_{in}$ of the inclusive residual distribution is typically smaller than the true intrinsic resolution because including the hit of the probed detector in the track fit biases the fit by pulling it towards the hit in the probed detector. 

Ref.~\cite{NIMA-538-2005} originally showed that the intrinsic spatial resolution of a probed detector can be approximated by the geometric mean 
\begin{equation}
 \sigma=\sqrt{\sigma_{ex}\cdot\sigma_{in}} \label{geometricmeaneq}
\end{equation}
of the two residual widths. Specifically, the appendix of that paper gives an analytical derivation of eq.~\ref{geometricmeaneq} that relies on an approximation for the expression of the variance of the residual distributions. The advantage of this simple method for analyzing tracking data from beam tests is that the resolutions for reference tracking detectors do not have to be known and no tracking simulations are required in order to estimate the resolution of a probed detector. 

However, Monte Carlo simulation studies later showed that this geometric-mean method only produces accurate results when the resolutions of all tracking detectors and of the probed detector are similar to each other~\cite{JINST-2014}. This makes the geometric-mean method applicable if, for example, tracking detectors and probed detector are all of the same type and constructed in the same way and hence can be expected to have similar resolutions. Ref.~\cite{JINST-2014} showed that if the resolution of a probed detector is significantly worse than the resolutions of the tracking detectors, e.g. 250~$\mu$m for the probed detector vs.\ 50~$\mu$m for the tracking detectors, then the resolution calculated with the geometric-mean method will be about 12\% better than the true intrinsic resolution of the probed detector. This might be the case, for example, when an MPGD is probed using tracks measured with a high-resolution silicon tracker. 

In this paper, we extend the study of the accuracy of the geometric-mean method to a more realistic situation in which multiple Coulomb scattering (MCS) of the tracked particles is also taken into account. This is done both with tracking data for GEM detectors from a beam test and with a stand-alone Geant4~\cite{Geant4Std} simulation created for the beam test data analysis.

\section{Detector configuration} \label{sec:config}
In the beam test we conducted, a total of ten triple-GEM detectors were operated as a tracking system~\cite{ZigzagStrip}. In this configuration, a total material amount corresponding to 14\% of a radiation length was placed in the beam, which produced non-negligible multiple Coulomb scattering of the tracked particles. The need to properly account for MCS effects in the spatial resolution measurements for the tracking detectors motivated the study presented here. In the test, four GEM detectors were used as reference tracking detectors and all six probed GEM detectors were placed between the second and the third reference tracking detectors. 

\begin{figure}[htb]
  \centering
  \includegraphics[width=8cm, height=4cm]{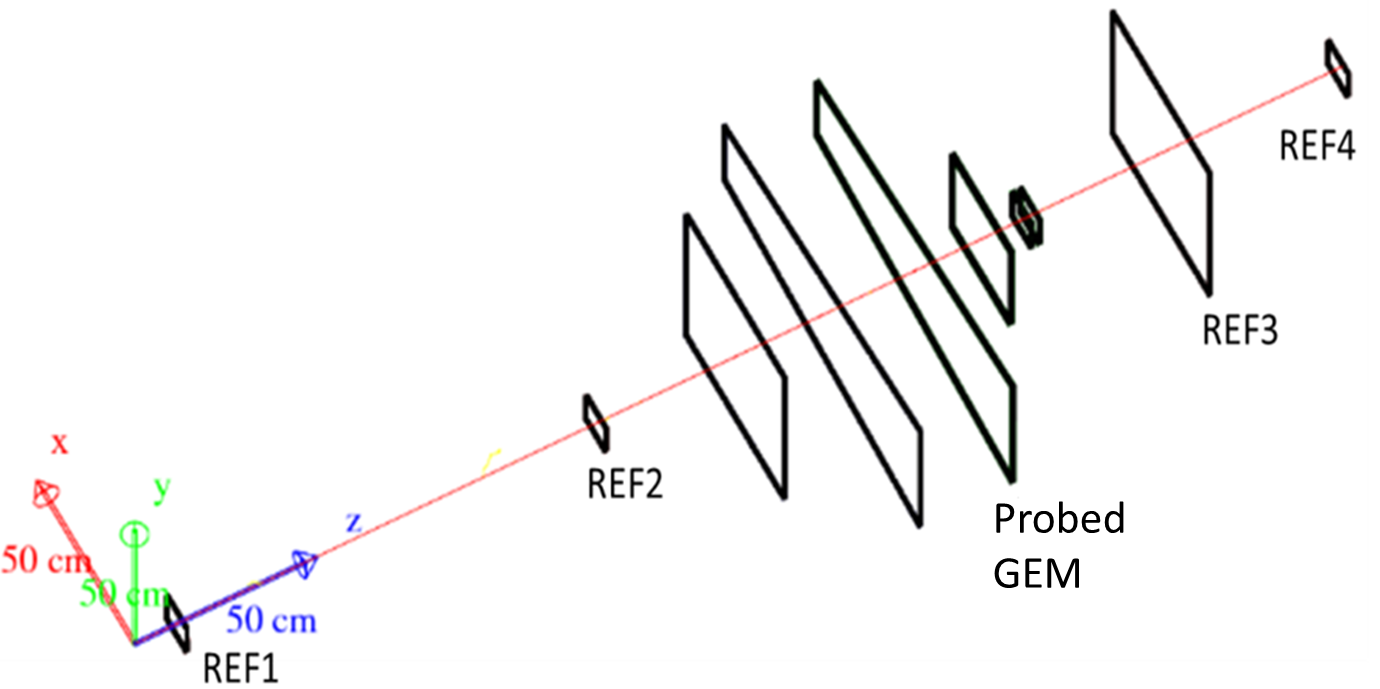}
  %\vspace{-0.5cm}
  \caption{Configuration of reference tracking detectors REF 1-4 and probed detector implemented in Geant4 for the beam test analysis.}
  \label{geometry}
\end{figure}

One of these probed GEM detectors is used for the study presented here. This detector could be equipped with two different readout boards that featured either zigzag strips~\cite{ZigzagStrip} or regular straight strips \cite{StraightStrip}. In order to extract the intrinsic resolutions for this GEM detector in those two configurations, a stand-alone Geant4 simulation was created in which MCS due to the materials of all ten detectors was taken into account~\cite{ZigzagStrip}. Figure~\ref{geometry} shows the detector configuration that was implemented in Geant4. The first two and the last two detectors served as reference tracking detectors (denoted as REF1 - REF4). They were constructed in the same way with 2D readout. The distance between REF1 and REF4 was about 3 m. The probed GEM detector was placed between REF2 and REF3. The study presented here focuses only on determining the resolutions for these five detectors. Except for their contribution to the multiple scattering of tracked particles, the other five GEM detectors are ignored.

\section{Analysis method and results} \label{sec:results}

\subsection{Resolutions for tracking detectors in the simulation}
As a first step, the resolutions of the four reference tracking detectors determined with the geometric-mean method are compared with the intrinsic resolutions that are fed as inputs into the Geant4 simulation. When studying resolutions for one tracking detector, exclusive tracks are reconstructed from the other three reference tracking detectors and inclusive tracks are reconstructed from all four tracking detectors. In other words, the ``probed'' detector here is one of the four tracking detectors and the other three tracking detectors work as reference detectors.

In the first such scenario, we assume that the three reference detectors have perfect intrinsic resolutions. The tracking detector under study is assigned a resolution from 10~$\mu m$ to 140~$\mu m$ in steps of 10~$\mu m$. Using the simulated track data, geometric means of inclusive and exclusive residual widths are calculated. 

We then attempt to statistically correct these geometric-mean resolutions for MCS effects. If the resolution of the probed detector is also set to be perfect in this simulation scenario, then any non-zero residual width is purely due to MCS in the material. For example, in our configuration the exclusive (inclusive) ``pure MCS residual'' in the x-coordinate has a width of 46~$\mu m$ (32~$\mu m$) for REF2. The geometric mean of these MCS residual widths is 38~$\mu m$. We subtract this value in quadrature from the corresponding geometric mean numbers for the REF2 detector obtained with non-perfect input resolutions as an attempt to correct the resolutions for MCS. Resolutions are calculated a third way by determining the track error at the position of the probed detector and then subtracting it in quadrature from the width of the exclusive residual. More details about this track-error estimate method can be found in \cite{ZigzagStrip}. 

\begin{figure}[htb]
  \begin{center}
      \vspace{-0.3cm}
        \subfigure[REF1X]{
          \label{TrackersNotSmear:REF1X}%%
          \begin{minipage}[b]{0.5\textwidth}
            \centering
            \includegraphics[width=7.5cm]{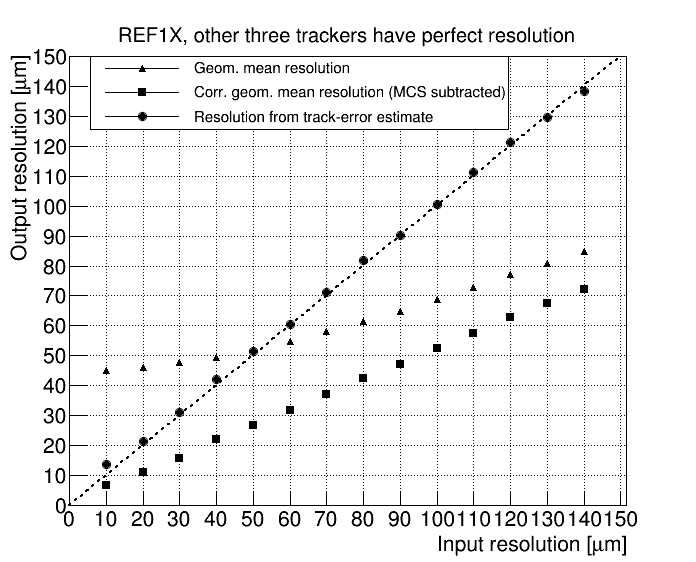}
          \end{minipage}}%
       %\vspace{-0.3cm}
        \subfigure[REF2X]{
          \label{TrackersNotSmear:REF2X}%%
          \begin{minipage}[b]{0.5\textwidth}
            \centering
            \includegraphics[width=7.5cm]{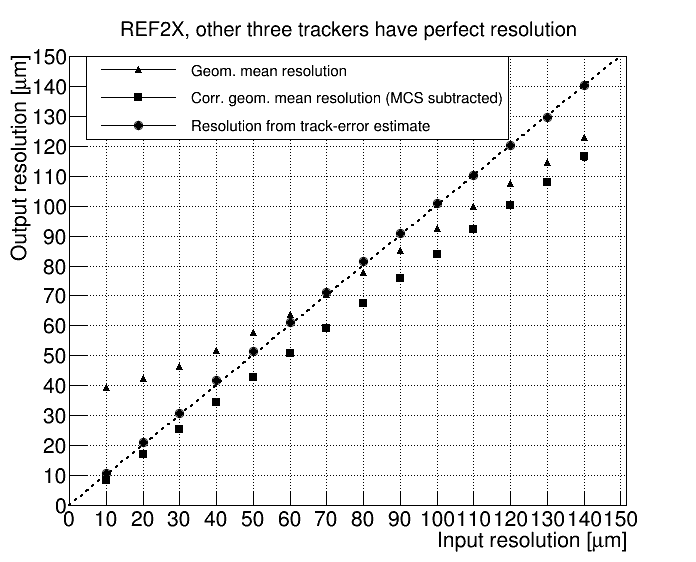}
          \end{minipage}}\newline%
      \vspace{-0.3cm}
        \subfigure[REF3X]{
          \label{TrackersNotSmear:REF3X}%%
          \begin{minipage}[b]{0.5\textwidth}
            \centering
            \includegraphics[width=7.5cm]{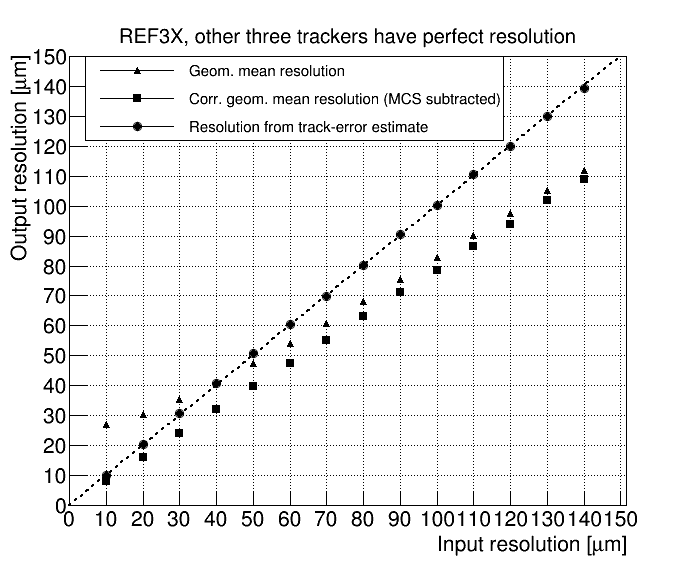}
          \end{minipage}}%
      % \vspace{-0.3cm}
        \subfigure[REF4X]{
          \label{TrackersNotSmear:REF4X}%%
          \begin{minipage}[b]{0.5\textwidth}
            \centering
            \includegraphics[width=7.5cm]{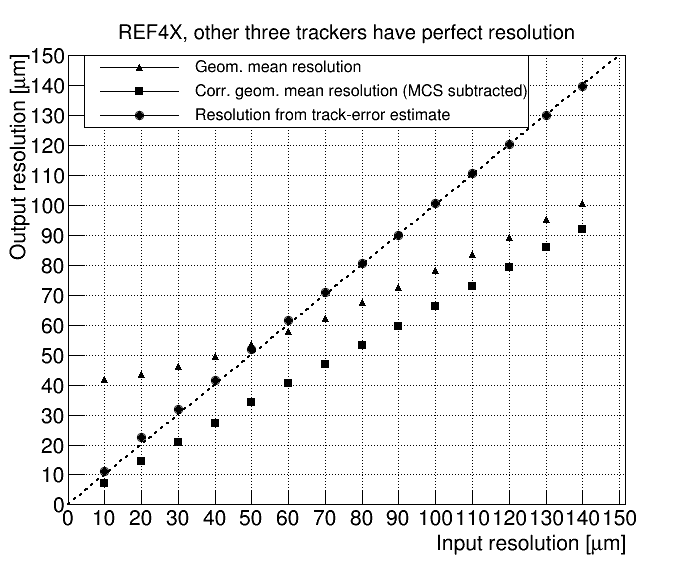}
          \end{minipage}}%%
        \vspace{-0.3cm}
        \caption{Resolutions for tracking detectors in x-coordinate measured with the geometric-mean method with and without subtracting the MCS effect vs.\ resolutions put into the Geant4 simulation. Here all three tracking detectors which are used to build tracks have perfect resolutions. Resolutions measured with the track error estimate are also shown. Statistical errors are smaller than marker sizes.}
        \label{TrackersNotSmear}
  \end{center}
\end{figure}

The accuracies of the three resulting resolutions (of the x-coordinate) are compared in figure~\ref{TrackersNotSmear} by plotting them against the input resolution. We find that the resolutions measured from the track error estimate correctly reproduce the input resolutions over the full range of inputs. Resolutions measured with the geometric-mean method are systematically biased giving a worse estimate for good resolutions (10-40~$\mu m$) and an overly optimistic estimate for worse resolutions (60-140~$\mu m$). Statistically correcting residuals for MCS effects as described above causes the estimate with the geometric-mean method to produce overly optimistic results over the entire studied range. This effect is more pronounced for the first and last detectors (REF1 and REF4) because in those two cases the reference tracks from the other three detectors are extrapolated rather than interpolated to the position of the probed detector. 

\begin{figure}[htb]
  \begin{center}
     \vspace{-0.3cm}
        \subfigure[REF1X]{
          \label{TrackersSmear50:REF1X}%%
          \begin{minipage}[b]{0.5\textwidth}
            \centering
            \includegraphics[width=7.5cm]{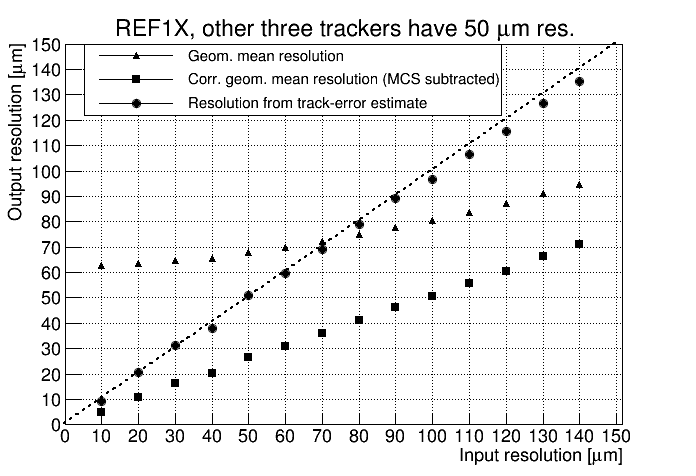}
          \end{minipage}}%
      % \vspace{-2cm}
        \subfigure[REF2X]{
          \label{TrackersSmear50:REF2X}%%
          \begin{minipage}[b]{0.5\textwidth}
            \centering
            \includegraphics[width=7.5cm]{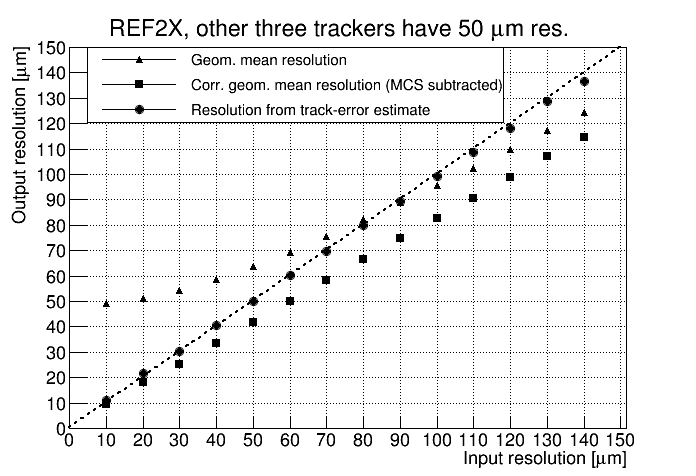}
          \end{minipage}}\newline%
      \vspace{-0.3cm}
        \subfigure[REF3X]{
          \label{TrackersSmear50:REF3X}%%
          \begin{minipage}[b]{0.5\textwidth}
            \centering
            \includegraphics[width=7.5cm]{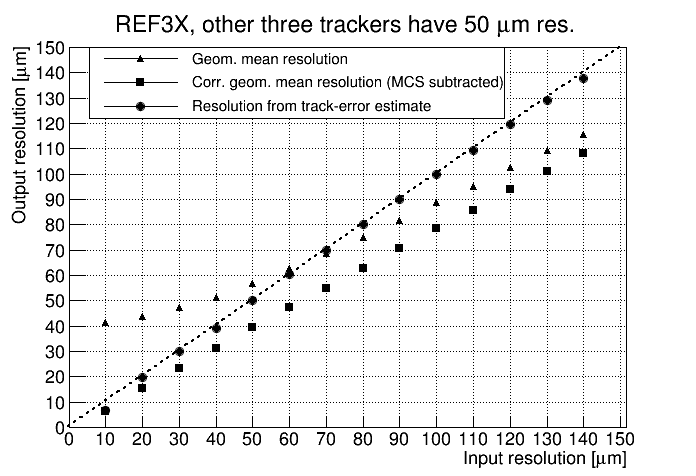}
          \end{minipage}}%
      % \vspace{-2cm}
        \subfigure[REF4X]{
          \label{TrackersSmear50:REF4X}%%
          \begin{minipage}[b]{0.5\textwidth}
            \centering
            \includegraphics[width=7.5cm]{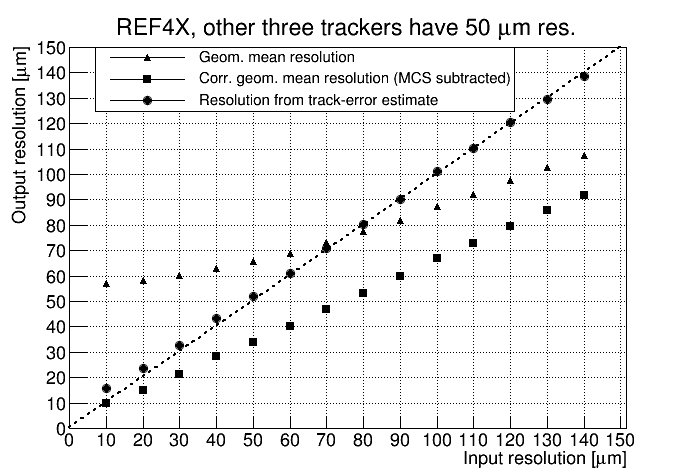}
          \end{minipage}}%%
        \vspace{-0.3cm}
        \caption{Resolutions for tracking detectors in x-coordinate measured with the geometric-mean method with and without subtracting the MCS effect vs.\ resolutions put into the Geant4 simulation. Here all three tracking detectors which are used to build tracks have 50~$\mu m$ resolution. Resolutions measured with track error estimate are also shown. Statistical errors are smaller than marker sizes.}
        \label{TrackersSmear50}
  \end{center}
\end{figure}

In the second scenario studied, the three tracking detectors that are used to reconstruct tracks are assumed to have finite and identical resolutions of $\sigma = 50\ \mu m$. We pick this value since it represents the best resolution that can be achieved with GEM detectors. Results are shown in figure~\ref{TrackersSmear50}. In this case, the systematic bias introduced by the geometric-mean method is slightly larger than for the case of perfect tracking resolutions across most of the studied range of resolutions. Statistically correcting residuals for MCS effects again causes the estimate with the geometric-mean method to produce overly optimistic results over the entire range. 

\begin{figure}[htb]
  \begin{center}
     \vspace{-0.3cm}
        \subfigure[REF1X]{
          \label{TrackersSmear:REF1X}%%
          \begin{minipage}[b]{0.5\textwidth}
            \centering
            \includegraphics[width=7.5cm]{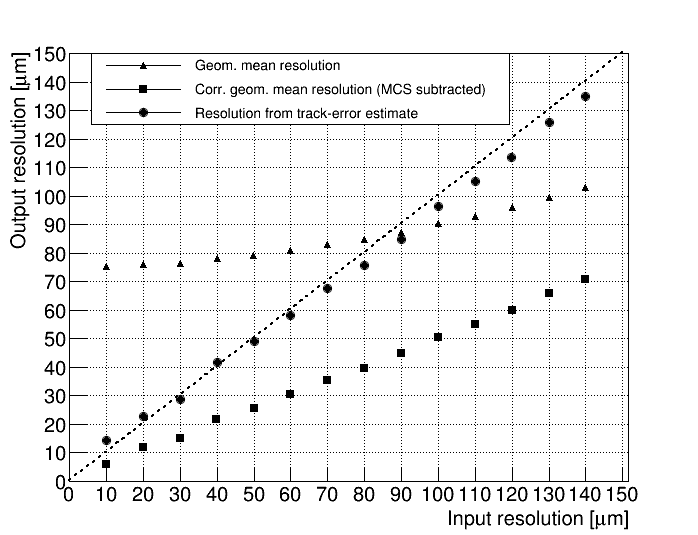}
          \end{minipage}}%
      % \vspace{-2cm}
        \subfigure[REF2X]{
          \label{TrackersSmear:REF2X}%%
          \begin{minipage}[b]{0.5\textwidth}
            \centering
            \includegraphics[width=7.5cm]{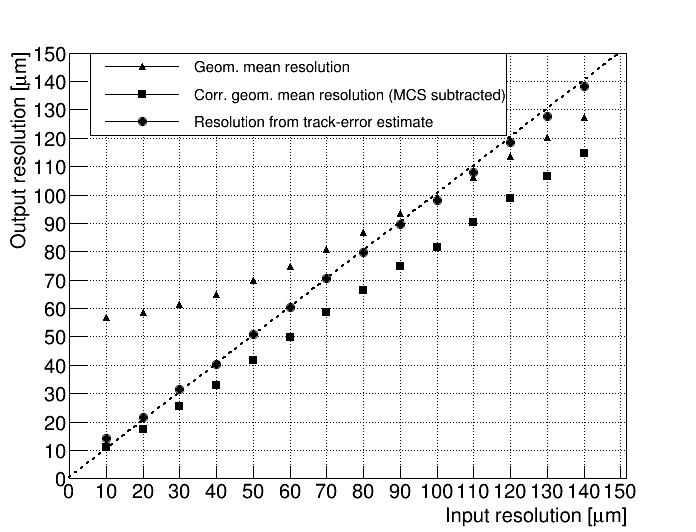}
          \end{minipage}}\newline%
     \vspace{-0.3cm}
        \subfigure[REF3X]{
          \label{TrackersSmear:REF3X}%%
          \begin{minipage}[b]{0.5\textwidth}
            \centering
            \includegraphics[width=7.5cm]{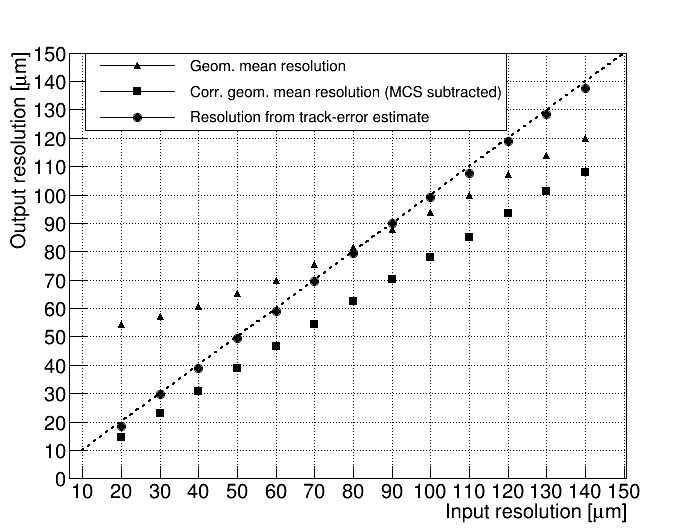}
          \end{minipage}}%
      % \vspace{-2cm}
        \subfigure[REF4X]{
          \label{TrackersSmear:REF4X}%%
          \begin{minipage}[b]{0.5\textwidth}
            \centering
            \includegraphics[width=7.5cm]{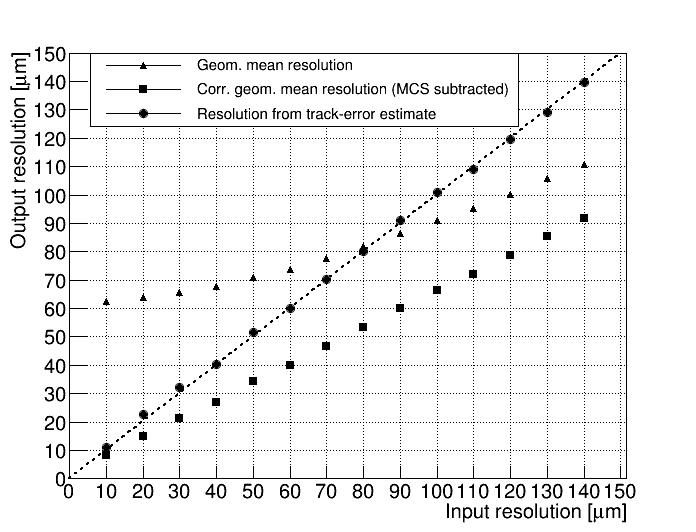}
          \end{minipage}}%%
        \vspace{-0.3cm}
        \caption{Resolutions for tracking detectors in x-coordinate measured with the geometric-mean method with and without subtracting the MCS effect vs.\ resolutions put into the Geant4 simulation. Here the three tracking detectors which are used to build tracks have realistic resolutions measured from the beam test data and they are slightly different from each other (see text). Resolutions measured with track error estimate are also shown. Statistical errors are smaller than marker sizes.}
        \label{TrackersSmear}
  \end{center}
\end{figure}

Figure~\ref{TrackersSmear} shows that this trend of increasing mismeasurement of resolutions continues when tracking detector resolutions measured with our actual beam test data, i.e.\ 73, 70, 59 and 68~$\mu m$~\cite{ZigzagStrip}, are used for REF1-4, respectively. We find that the mismeasurement of the intrinsic resolution of the probed detector can then be as large as 50\%. 

\subsection{Resolutions for the probed GEM detector in the simulation}
\begin{figure}[htb]
  \begin{center}
       \vspace{-0.3cm}
          \begin{minipage}[b]{0.5\textwidth}
            \centering
            \includegraphics[width=7cm]{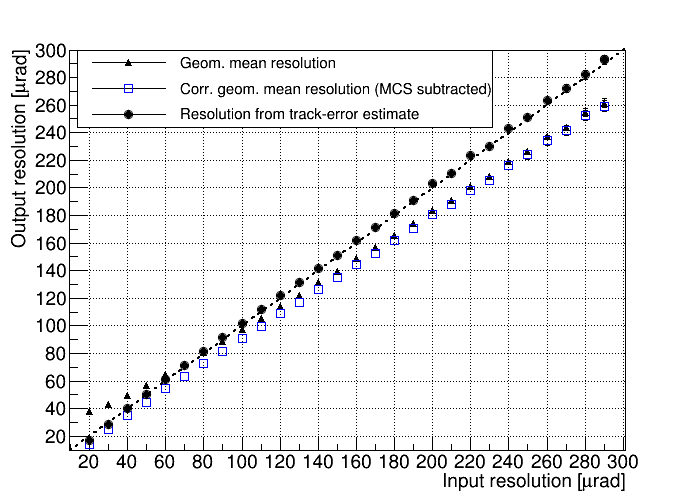}
          \end{minipage}%
      % \vspace{-2cm}
          \begin{minipage}[b]{0.5\textwidth}
            \centering
            \includegraphics[width=7cm]{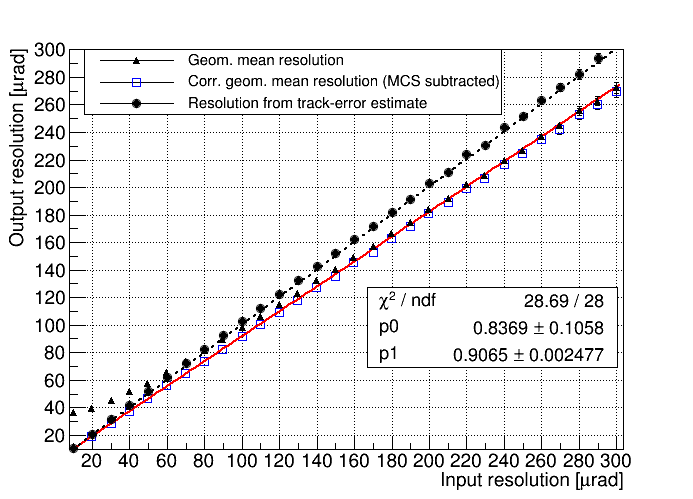}
          \end{minipage}%
        \vspace{-0.3cm}
        \caption{Resolutions obtained for the probed GEM detector in polar coordinate $\phi$ vs.\ resolutions put into the Geant4 simulation. Results are shown for the geometric-mean method with and without MCS effect subtracted and for the track error method. On the left, input resolutions for the four tracking detectors are set to 50~$\mu m$; on the right, actual resolutions for tracking detectors measured from the beam test data are used. Statistical errors are smaller than marker sizes.}
        \label{ProbedDetector}
  \end{center}
\end{figure}

Using the same method as in the previous section, the resolutions for the probed detector (see figure~\ref{geometry}) are studied in the Geant4 simulation. Here we study two cases with resolutions for all four tracking detectors either set to 50~$\mu m$ or to the realistic resolutions obtained from the beam test data as mentioned above. In this case, resolutions are studied in the $\phi$ coordinate of a polar coordinate system since the probed detector tested in the beam features radial readout strips and measures only $\phi$ coordinates. The transfer to the polar coordinate system and the relative alignment of the detectors in that system are detailed in Ref.\cite{ZigzagStrip}. The results in figure~\ref{ProbedDetector} show that the geometric-mean method mismeasures the resolution of the probed detector in a simular fashion as demonstrated in the previous section while the method based on track-error estimates again delivers accurate results. One difference from the previous section is that now the statistical subtraction of the MCS effect does not affect the results of the geometric-mean method very much above 100~$\mu m$. In that particular range, the geometric-mean resolution returns resolutions that are about 10\% too optimistic.

\begin{figure}[htb]
  \begin{center}
      \vspace{-0.5cm}
          \begin{minipage}[b]{0.5\textwidth}
            \centering
            \includegraphics[width=7cm]{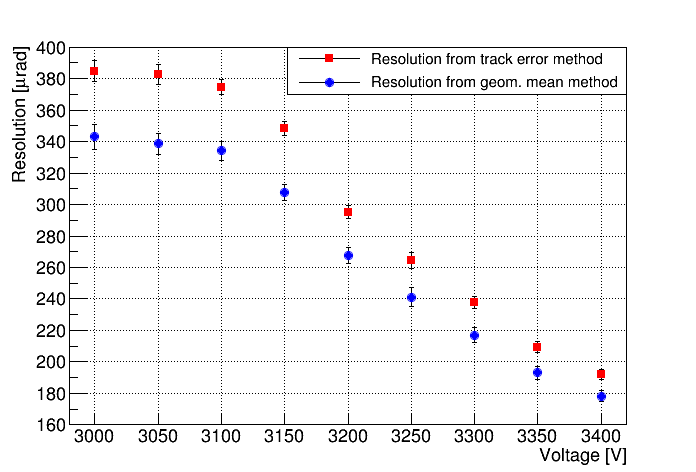}
          \end{minipage}%
      % \vspace{-2cm}
          \begin{minipage}[b]{0.5\textwidth}
            \centering
            \includegraphics[width=7cm]{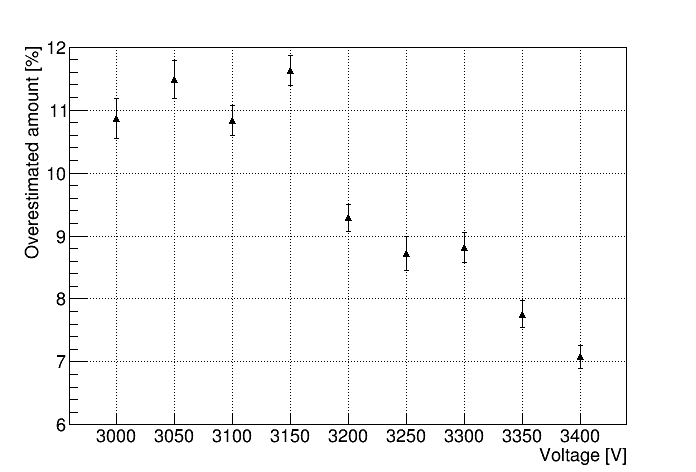}
          \end{minipage}%
        \vspace{-0.3cm}
        \caption{Left: Comparison of geometric-mean resolutions and track-error resolutions for the probed GEM detector read out with zigzag strips in polar coordinate $\phi$ vs.\ high voltage applied to the drift electrode of the detector. Right: Percent difference between track-error resolution and geometric-mean resolution at each voltage point.}
        \label{ProbedDetectorCom1}
  \end{center}
\end{figure}

\begin{figure}[htb]
  \begin{center}
      \vspace{-0.5cm}
          \begin{minipage}[b]{0.5\textwidth}
            \centering
            \includegraphics[width=7cm]{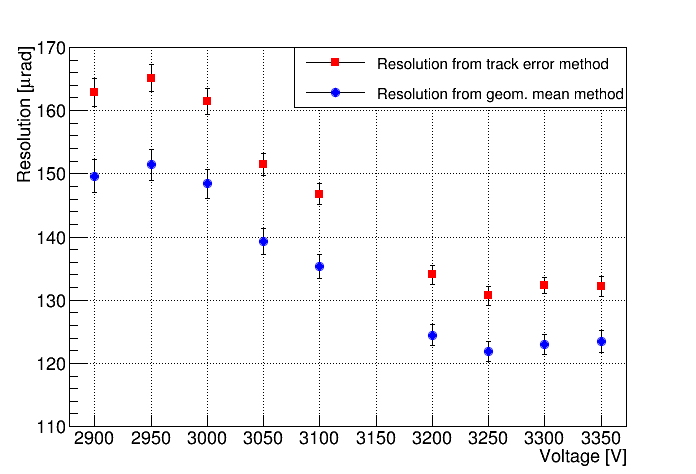}
          \end{minipage}%
      % \vspace{-2cm}
          \begin{minipage}[b]{0.5\textwidth}
            \centering
            \includegraphics[width=7cm]{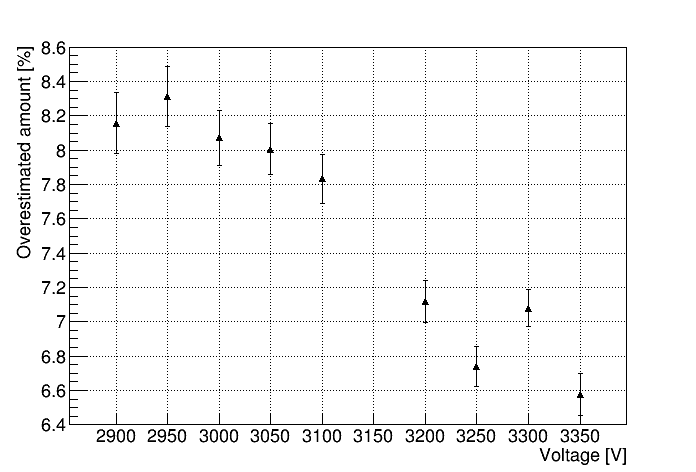}
          \end{minipage}%
        \vspace{-0.3cm}
        \caption{Left: Comparison of geometric-mean resolutions and track-error resolutions for the probed GEM detector read out with straight strips in polar coordinate $\phi$ vs.\ high voltage applied to the drift electrode of the detector. Right: Percent difference between track-error resolution and geometric-mean resolution at each voltage point.}
        \label{ProbedDetectorCom2}
  \end{center}
\end{figure}

\subsection{Resolutions for the probed GEM detector measured in beam test}
We apply both track-error method and geometric-mean method with MCS subtraction to beam test data collected with the probed GEM detector which is read out either with zigzag strips or with straight strips. Figures \ref{ProbedDetectorCom1} and \ref{ProbedDetectorCom2} show the resulting resolutions as a function of high voltage applied to the drift electrode of the detector. For zigzag strips, the resolution is in a range of 190-380~$\mu$rad; the geometric-mean method delivers numbers that are 7-11\% more optimistic than the track-error result~(figure \ref{ProbedDetectorCom1}). For straight strips, the resolution varies from 130 to 160~$\mu$rad and the geometric-mean method is 7-8\% too optimistic~(figure \ref{ProbedDetectorCom2}). The inaccuracy of the geometric-mean method observed with measured data is comparable to that found in the simulation.

\section{Summary and Conclusion}
Using both simulation and direct measurement, we have demonstrated that the spatial resolution of a position-sensitive detector obtained with the geometric-mean method is systematically biased also when non-negligible multiple Coulomb scattering must be taken into account. This extends previous studies that did not consider the impact of the scattering on the bias. Subtracting scattering effects in quadrature from resolutions obtained with the geometric-mean method typically leads to resolutions that are too optimistic by 10-50\%. The size of the bias that the method introduces depends on the number of reference detectors and on the location of the probed detector within the tracker.

In conclusion, the geometric-mean method can be used for rough estimates of the resolution of a tracking detector without knowing much detail about the resolutions of the reference detectors in a beam test even if MCS is present. However, it is not an optimal method for measuring the intrinsic spatial resolution of a detector due to a lack of accuracy. To obtain accurate results the track-error method is recommended instead.

\acknowledgments
This work is supported by Brookhaven National Laboratory~(BNL) under the EIC eRD-6 consortium. The authors would like to thank Alexander Kiselev (BNL) and Kondo Gnanvo (U. of Virginia) for their very helpful discussions and suggestions on the analysis methods.

\end{document}